\documentclass[conference,10pt]{IEEEtran}

\usepackage{times}
\usepackage{amsmath}
\usepackage{amssymb}
\usepackage{algorithm}
\usepackage{algorithmic}
\usepackage{color, array, colortbl}
\usepackage{graphicx}
\usepackage{multirow}
\usepackage{cancel}
\usepackage{mdwlist}

\usepackage{framed,color}
\definecolor{shadecolor}{rgb}{0.9,0.9,0.9}

\usepackage{algorithm}
\usepackage{algorithmic}

\newcommand{\CloudNet}{\text{CloudNet}}
\newcommand{\CloudNets}{\text{CloudNets}}

\newcommand{\mrm}{\mathrm}

\newcommand{\ignore}[1]{}

\begin{document}






\title{Generalized and Resource-Efficient VNet Embeddings with Migrations}

\author{
Gregor Schaffrath, Stefan Schmid, Anja Feldmann\\
Telekom Innovation Labs \& TU Berlin, Berlin, Germany \\
\{grsch,stefan,anja\}@net.t-labs.tu-berlin.de
}


\maketitle

\sloppy


\begin{abstract}
This paper attends to the problem of embedding flexibly specified \CloudNets, virtual networks connecting cloud resources (such as storage or computation). We attend to a scenario where customers can request \CloudNets\ at short notice, and an infrastructure provider (or a potential itermediate broker or reseller) first embeds the \CloudNet\ fast (e.g., using a simple heuristic). Later, however, long-lived \CloudNets\ embeddings are optimized by migrating them to more suitable locations, whose precise definition depends on a given objective function. For instance, such migrations can be useful to reduce the peak resource loads in the network by spreading \CloudNets\ across the infrastructure, to save energy by moving \CloudNets\ together and switching off unused components, or for maintenance purposes.

We present a very generic algorithm to compute optimal
embeddings of \CloudNets: It allows for different objective
functions (such as load minimization or energy conservation),
supports cost-aware migration, and can deal with all link types that
arise in practice (e.g., full-duplex or even wireless or wired
broadcast links with multiple endpoints). Our evaluation shows that such a rigorous optimization is even feasible in order to optimize a moderate-size \CloudNet\ of full flexibility (e.g., a router site, a small physical infrastructure or virtual provider network).
\end{abstract}

\section{Introduction}
\label{sec_intro}

More and more of today's infrastructure is being virtualized.
Emerging link virtualization technologies such as OpenFlow allow us
to realize the vision of \CloudNets\ which provide an abstraction of
\emph{both nodes and links}, connecting (and providing access to) virtual cloud resources with virtual networking.
Decoupling virtual networks from the
physical constraints of the underlying infrastructure (the
\emph{substrate}), \CloudNets\ can offer opportunities for
customized network environments and can be flexibly \emph{embedded}\footnote{In this paper, the terms
embedding and mapping are treated as synonymous and used interchangeably.} at
optimal (e.g., economical) locations and even migrated.

One of the central challenges arising from \CloudNets\ concerns the
strategies to leverage the resource allocation flexibility. While
previous work focused on optimized initial embeddings of graph-like
topologies~\cite{infocom2009}, recent efforts aim at the improvement
of the migration process~\cite{clnetvee11} or the calculation of the
most beneficial reconfigurations under constraints of additional
network communication cost~\cite{clwardsigc11}. However, the
question of how virtual network embedding costs are affected by the
possibility of migration has received much less attention so far. While
today's network virtualization technology facilitates seamless
migration (without session interruption), migration inevitably
introduces costs (e.g., in terms of computation, bandwidth, or even
roaming fees in case of cross-provider migrations). Whether and
where to migrate (parts of) virtual networks is hence a non-trivial
problem.

Our work is based on (and incorporated in) our prototype
virtualization architecture described in~\cite{visa09virtu}. The
architecture is motivated by both technical and business
perspectives. It considers a scenario where participating entities
focus on abstractions relevant to them and optimize towards their
own goals: A customer requests a network with (possibly incomplete)
requirement specifications, whereas every unspecified parameter is a
degree of freedom to the providers. Brokers and resellers without
holistic substrate topology knowledge split up the request and
negotiate for partial resources with cloud providers, optimizing the
embedding towards their own benefits. Cloud providers in turn
optimize the embeddings inside their own substrate segment along
their specific substrate management policies. We further assume that
large cloud providers may substructure into sites, and introduce
their own internal reseller or broker instances.

This scenario entails two important implications: First, the
abstraction characteristic of virtualization and the given
requirement specification allow providers on any level (cloud
providers and resellers) to freely optimize embeddings for their own
purposes. If agreed upon provisioning properties are met, changes
may be assumed not to disrupt applications inside the \CloudNet.
Second, resource mapping---a computationally hard problem---needs to
scale only to the scope of one role/player, as it happens in
multiple steps.

The embedding strategies and objectives are likely to vary in size
and heterogeneity of the available substrate, as well as the time
available to find a solution. It is unlikely that one \CloudNets\
embedding approach will fit all situations: Symmetries in data
centers may allow for reasonable approximations e.g., by greedy
virtual network placements or aggregation of similar resources. A
backbone router site may be too diverse for such an approach, but
allow for the computation of optimal solutions to its limited size.
In scenarios with heavy-tailed \CloudNet\ durations, short term
placements may not even warrant the effort of optimization, but it
is beneficial for a provider to optimize \emph{long-lived} \CloudNets\
into an efficiently managed segment of the network.

\subsection{Our Contribution}

This paper addresses the \CloudNets\ embedding problem. We propose a
generic embedding algorithm which \emph{jointly optimizes node
placements and link embeddings} and exploits this flexibility to
compute \emph{optimal embeddings} (``realizations'') of the
\CloudNets, while \emph{guaranteeing} the allocation and
provisioning of the requested combination of resources. Our
algorithm is the first one to integrate \emph{cost-aware
migrations}. We believe that as the type and arrival time of
\CloudNet\ requests is hard to predict, the possibility of
reconfigurations and migration is crucial. Moreover, our algorithm
does not rely on any particular ``clean-slate'' type of substrate
topology~\cite{rethinking}.

In addition, our approach provides high flexibility, including all
link types that occur in practice, such as half-duplex, full-duplex,
or even broadcast links with multiple endpoints (as they appear in
wireless networks but also in wired contexts with hubs), in the
sense that any of these links can be mapped on any other. It
supports embeddings across resources and resource types and exact
solutions for the mapping of partial networks to single substrate
nodes.
It further supports provider-side placement policies as well as
resource prioritization (e.g., prioritizing lucrative resource
allocations). It also allows us to take into account node-based
loads, e.g., as a function of the packet rate (shorter packets
increase the computational load at the forwarding engine) etc.

Interestingly, despite this flexibility, our algorithm is a (linear)
Mixed Integer Program (MIP) and can hence be solved by standard and
optimized tools such as \texttt{CPLEX}. Another advantage of the
mathematical programming approach is that it enables us to propose
different objective functions which can be easily exchanged. For
example, at some point a provider may choose to place the virtual
networks on the ``edge'' of the physical network in order to avoid
blocking bottleneck links and hence to maximize the likelihood that
future \CloudNet\ requests can be accepted. At another point it
wants to spread the \CloudNet\ embeddings as much as possible in
order to minimize the load or congestion, or to collocate the
\CloudNets\ as much as possible in order to be able to switch off
other parts of the network in order to save energy or for
maintenance work.

In contrast to various existing embedding heuristics for
virtual networks (e.g.,~\cite{ammar,pb-embed,turner}), the focus of
this paper is on an \emph{integrated} approach which emphasizes the
\emph{quality} of the \CloudNet\ embeddings. In
particular, we focus on the optimal solutions, and investigate the
feasibility of this approach to improve initially heuristically
placed \CloudNets; this is interesting for \emph{long-lived} \CloudNets\ where
the resource investments for computing the optimizations may pay off in the future.
We have three use-cases in mind: (1) a
\emph{VPN-like scenario} where the virtual node locations are given,
(2) a \emph{data center like scenario} where the virtual node
placements are fully flexible, and (3) an \emph{out-sourcing / cloud
scenario} where some virtual nodes have a fixed location (location
and access network of a company) and others do not (out-sourced
services, to cloud). Clearly, the data center scenario is the most challenging
for optimization, while the VPN scenario, without the possibility to optimize terminal
mappings, boils down to a classic flow problem which can even be solved optimally in polynomial time. In our
evaluation, we will hence focus on the first two.

Note that our algorithm allows us to compute, e.g., the migration
cost-benefit tradeoff: By computing the embedding that would result
from migration together with the migration cost, it is left to the
(potentially automated) administrator to decide whether the changes
are worthwhile. For example, our algorithm allows to answer
questions such as: Can we migrate \CloudNets\ to a more compact form
such that 20\% of the currently used resources are freed up, and
what would be the corresponding migration cost?

We have implemented all our algorithms in our \CloudNets\ prototype
architecture.

%

\section{Challenges}\label{sec:challenges}

A distinguishing feature of \CloudNets\ is the flexibility in terms
of the specification and the combination of different resources,
from bandwidth requirements along virtual cloud links to storage and
computational capacities at the virtual cloud nodes. Indeed, we
consider all kinds of real live constraints. This requires a generic
and formal interface that allows a customer and a provider to agree
on a certain service; within this specification, the provider is
free to optimize and re-embed the allocation, e.g., in order to make
optimal (re-)use of the given infrastructure, to provide the best
service or to save energy by switching off unused resources.
The
specification of a \CloudNet\ can include capacities, geographical
constraints, specific versions of the operating system, or also
non-topological requirements such as the binary compatibility (e.g.,
w.r.t.~word size) of the architectures where two virtual nodes are
mapped to.

There are also constraints on the physical infrastructure provider
side. For example, not all possible link types are available, and
hence a virtual broadcast link may be implemented with multiple
asymmetric links. Moreover, a provider may have its own set of
policies where to map certain components.

Efficient \CloudNet\ embeddings are challenging and due to the large
number of possible specifications, a very flexible algorithm is
needed.
An algorithm to compute a good embedding of a given static set of
\CloudNets\ is not satisfactory in practice, as \CloudNet\ requests
arrive over time and an optimal placement at some time $t_0$ may
become suboptimal at some time $t>t_0$. For instance, if another
virtual cloud network expires, resources are freed up and another
\CloudNet\ could be migrated there. Re-embeddings can also make
sense for network management and maintenance, e.g., to move the
traffic to different paths to upgrade the routers. Moreover, the
demand for a certain service can be \emph{dynamic}, due to daytime
reasons, and also the origins of the request can change due to
\emph{user mobility} or \emph{time-zone effects}, which means that
\CloudNets\ should be dynamically scaled up or down depending on the
demand, and moved with the users to ensure a good latency of the
access (Quality-of-Service/Quality-of-Experience parameters). An
embedding algorithm must hence support cost-aware migrations in the
sense that it trades off migration cost against the potential
benefits.

%
%
%

\subsection{Migration Cost}\label{sec:map:cost}

While migrations may yield more efficient embeddings, their costs depend on many factors. For instance, classic server migrations may entail service interruption costs that depend on the available bandwidth along the migration path, while live migration technology may provide a seamless service and only come at a cost of bulk data transfers. In this paper, we will group these factors into three categories:

\noindent\textbf{Resource removal:} These costs are independent of the migration destination. They either relate to the removal of old allocations, or to the fact that a migration is happening per se. A migration entails a management overhead $C_{\mrm{mgmt}}$. If, e.g., a virtual network provider triggers a cross-provider migration, the termination of provisioning contracts may entail penalties $C_{\mrm{contract}}$.
		Temporary redundant allocations of resources and reconfiguration based service outages during migration entail opportunistic costs $C_{\mrm{reconfig}}$.
		An example cause for the latter would be outages triggered by switchovers to and from transitional provisioning solutions.

\noindent \textbf{Resource transferral:} These costs may depend on the migration destination. They relate to the actual transfer and possible property changes.
		Bulk data transfers may entail both real and opportunistic transit costs $C_{\mrm{transit}}$.
		E.g., transfer of host state may require additional bandwidth to be leased from transit providers. It may interfere with provisioning of other \CloudNets, if routed via the same substrate links.
		Furthermore, adaptations may impose overhead if crucial properties change (e.g., migrating a virtual host from Xen to KVM). We denote this cost factor by $C_{\mrm{adaptation}}$.

\noindent \textbf{Resource reinstantiation:} Establishment of new provisioning contracts or cost and benefit changes relate to the new placement and hence the future position.
		However, we model them in the context of placement preferences, as they are semantically equivalent to the factors influencing initial placement.
		This becomes clear considering that some resources (e.g., virtual links) may be migrated by reinstantiation rather than actual transfer.

Some of the mentioned costs may be zero and others approximated for practical reasons. As an example, consider the following two scenarios:
The first comprises a live migration of a host inside of a (fully switched, homogeneous) rack belonging to the same provider, providing separate links for the data plane and migrations. Evidently, contract penalties do not apply, no adaptations are required, and transit costs may be approximated as destination independent in the scope.
In the second, a virtual network provider live-migrates a host between providers. All cost factors may apply, but those of the second group may be included in $C_{\mrm{reconfig}}$, if the migration is provided as a service on behalf of the physical infrastructure providers.

\subsection{Use Cases}

We have three basic use cases in mind.

\noindent \textbf{VPN:} In the \emph{VPN scenario}, the locations of the virtual nodes
of a \CloudNet\ are fully specified. In contrast to typical VPN
networks however, resources are reserved along the paths connecting
the VPN terminals (admission control and traffic shaping to ensure
QoS), which essentially boils down to a network flow problem;
moreover, computational and storage resources can be specified at
the terminals.

\noindent \textbf{DC:} The \emph{data center scenario} describes the other
extreme where the virtual cloud nodes have full placement
flexibility and can be mapped to arbitrary locations.

\noindent \textbf{OC:} The so-called \emph{out-sourcing/cloud scenario} is situated between
the two extremes modeled by the static VPN and the fully flexible
DC scenario. A \CloudNet\ consists of some virtual nodes with fixed
locations (e.g., branches of the company and access network) while
other virtual nodes (providing, for example, storage or computation)
are ``out-sourced'', e.g., to a data center, and have a flexible
location.

Note that the use cases are related, and the VPN problem is often a
subproblem of the OC problem. An OC problem can also generate a DC
problem in the sense that once an appropriate data center is found,
an DC problem needs to be solved within the data center.

\section{Embedding: Key Concepts}\label{sec:setup}

The main objective of our algorithm is to embed \CloudNet\ requests
that arrive over time by mapping them onto the given substrate
network resources in such a way that the specification is fulfilled
(i.e., all specified resources are allocated to the \CloudNet); or
to reject the request otherwise. For instance, a virtual node may
require a 1GHz CPU and may only be mapped onto Linux nodes in the
US. Similarly, \CloudNet\ links may need a minimum of 10 MBit/s. The
resources offered by the substrate nodes and links can be shared
among the virtual networks.

The algorithm should be flexible in terms of the objective
function to be optimized for the \CloudNet\ placement, and the
arrival of a new \CloudNet\ should cause minimal changes to the
existing embedding of prior \CloudNets\ (i.e., since \CloudNet\
migration cost is non-zero there
  is a tradeoff between migration cost and a superior embedding).

In the following, we introduce the main ideas of our embedding
algorithm. We pursue a mathematical programming approach and present
a (linear) \emph{Mixed Integer Program (MIP)} which has the
advantage that standard software tools such as \texttt{CPLEX} or
\texttt{lpsolve} can be used to perform the computations. (Although
we focus on optimal embeddings, these tools also offer different
heuristics for faster but approximate solutions; all these
heuristics are directly applicable to our program as well.)

A MIP consists of a (linear) \emph{objective function} expressed
using a set of variables, plus a set of (linear) constraints on
these variables that ensure ``valid'' solutions. If a problem can be
specified in this form (what we do in this paper for the \CloudNet\
embedding problem), state-of-the-art optimized algorithms can be
used for evaluation. This section serves to introduce the reader to
our approach and the different variables and constants used in our
program. The complete formal program description appears in
Figure~\ref{fig:constraints}.

\subsection{Graph Representation}\label{ssec:defs}


\emph{Shared communication channels}, i.e., links with several end
points (both in the virtual and the substrate network) constitute a
first challenge for such a generic mapping approach.
To describe virtual and substrate networks as classic graphs $G=(V,E)$
consisting of vertices $V$ that are connected pairwise by edges $E$,
we introduce the notion of \emph{network elements} (NEs):
network elements represent both \emph{nodes} (set $NE_\mrm{N}$) and \emph{links}
(set $NE_\mrm{L}$). Network elements are connected by interfaces, which
form the edges of the graph.

We distinguish between virtual network elements of the \CloudNet\
(set $NE_\mrm{V}=NE_{\mrm{VN}}\cup NE_{\mrm{VL}}$ of virtual nodes and virtual links)
and substrate network elements of the substrate network (set
$NE_\mrm{S}=NE_{\mrm{SN}}\cup NE_{\mrm{SL}}$). In principle, any virtual node can be
mapped onto any substrate node, depending on the
requirements. A virtual link can be embedded onto a substrate node,
a substrate link, or onto a set of paths in the substrate network
(resulting in a multi-flow embedding).


The purpose of the embedding algorithm is to find a mapping of the
virtual networks and their elements to the network elements of the
substrate. To handle links with several endpoints, we replace
each link with a vertex and add graph edges accordingly.

\subsection{Placement Policies and Suitability} \label{ssec:map:approach:policies}

We use the binary matrix $new(u,v)$ to denote whether a virtual
cloud network element $u\in NE_\mrm{V}$
is mapped to a substrate network element $v \in
NE_\mrm{S}$ ($new(u,v)=1$) or not ($new(u,v)=0$).

A substrate element allocates resources for all virtual elements it
hosts. To describe these allocations, we introduce the variables
$alloc_{r_\mrm{V}}(u,v)$ which captures the amount of virtual resource $r_\mrm{V}$ of $u$ hosted on $v$ and
$alloc_{r_\mrm{S}}(u,v,r_\mrm{V})$ describing
the substrate resources $r_\mrm{S}$ used to allocate it.
The resources $r_\mrm{V}$ requested for $u$
are represented by the constant matrix $req(u,r_\mrm{V},s)$, where $s\in VT$ refers to the value type of request (e.g., minimum, maximum, ...).
To ensure that
the sum of the allocated resources never exceeds substrate
capacities of substrate we use the constant capacity matrices $cap_{r_\mrm{S}}(v)$, $cap_{r_\mrm{S}}(v,w)$,
and $cap(r_\mrm{S})$. The first two hold individual capacities of substrate components $v$ and substrate interfaces interconnecting $v$ and $w$ with respect to $r_\mrm{S}$. The last represents the capacity of a resource $r_\mrm{S}$ itself.
All three are required to correctly model various possible shared resources assignments in the substrate.

It is not always possible to map a virtual network element to any
arbitrary substrate element. For example, a virtual cloud node may
be restricted to substrate elements within the US. The constant
binary matrix $suit(u,v)$ specifies whether $v$ is suitable to host
network element $u$ ($suit(u,v)=1$) or not.

Our mathematical program considers placement restrictions: a
provider may want to bias or fix a mapping for a specific \CloudNet\
according to internal placement policies or cost factors. We thus use a constant
weight matrix $weight(u,v)$ to introduce a cost for each node
placement. These weights can also be used as policy support to prioritize certain
resource allocations over others in the objective function.

\subsection{Link Types and Resources} \label{ssec:map:approach:links}

Next, we discuss how we handle the different link types: If the
bandwidth in both directions is the same we call a link
\emph{symmetric}, otherwise it is called \emph{asymmetric}.  A
\emph{full-duplex} link supports traffic in both directions
\emph{independently}. A full-duplex link can be regarded as two
independent \emph{unidirectional} links. A shared (wireless, or
non-switched, hub-like) channel is referred as \emph{half-duplex}
link. Note that half-duplex links are symmetric by nature.

We explicitly distinguish between two classes
of resources $R=R_\mrm{V}\cup R_\mrm{S}$, namely virtual
resources $r_\mrm{V}\in R_\mrm{V}$ and substrate resources
$r_\mrm{S} \in R_\mrm{S}$.
To handle the different link types, virtual half-duplex links are associated to an $r_\mrm{V}$ of attribute
$\texttt{'/link}$$\texttt{/symmetric}$$\texttt{/bandwidth'}$
whereas substrate full-duplex links receive two $r_\mrm{S}$ with
$\texttt{'/link}$$\texttt{/upstream}$$\texttt{/bandwidth'}$ and
$\texttt{'/link}$$\texttt{/downstream}$$\texttt{/bandwidth'}$ respectively.
In our embedding program, we
assume a proportional relationship between $r_\mrm{V}$ and $r_\mrm{S}$, that is,
we consider a proportional factor $prop(r_\mrm{V},r_\mrm{S})$.
As different relation functions are possible (e.g., involving constant instantiation overhead), the respective constraints should be considered exemplary.

Interestingly, differentiating between $r_\mrm{V}$ and $r_\mrm{S}$ in both
\CloudNet\ specification and MIP is not only useful for
handling different link types but also for mapping nodes:
It enables us to map and even split resources of arbitrary resource types onto arbitrary other resource types.

To handle shared communication
channels we decompose its multiple endpoints into a \emph{set of
flows}. In particular, for each link $u$, we introduce a set $Fl(u)$
that describes the set of possible source-sink pairs for $u$.

Each flow $f\in Fl(u)$ inherits the requirements of $u$.
Analogously to the $alloc$ matrices, $flow_{r_\mrm{V}}(f,v,w)$ and $flow_{r_\mrm{S}}(f,v,w,r_\mrm{V})$ reflect tentative resource allocations on substrate interfaces,
and $new(f,v)$ denotes corresponding tentative flow mappings. Resources of these flows $f\in Fl(u)$ form the set $R_f \subset R_\mrm{V}$.

\subsection{The Flow Problem} \label{ssec:map:approach:flow}

While we consider virtual nodes atomic in the context of our MIP,
virtual links can be realized either as
single path or multiple paths within the substrate network. The
aggregated resources of the paths must satisfy the requirements of
the virtual link while not exceeding the capacity limits of the
substrate elements. For instance, the sum of the bandwidths of the
different paths must equal the link's bandwidth demand. This
constitutes a \emph{flow problem}. However, since we tackle
placement and embedding at the same time this corresponds to a
multi-commodity flow problem with a twist: The endpoints are not
fixed, but candidate locations overlap.

Our mathematical program ensures that the allocated flows are
connected, consistent with the requirements and capacities.
We enforce a \emph{flow
preservation invariant}, that is, we guarantee that the amount of
flow arriving at a node equals the amount of flow leaving the node.
However, we must exempt the source and the sink of the flow from
this invariant: We ensure that the traffic leaving the source equals
the demand of the virtual link. The link's sink simply consumes the
incoming flows. This is implemented via selector variables that
render the constraint trivially true for endpoints (a tautology).

\subsection{Migration Support} \label{ssec:map:approach:mig}

As mentioned, \CloudNet\ requests typically arrive over time and the provider
faces the problem of how to embed a new \CloudNet\ given the
existing allocations of other requests. Clearly, a complete
re-embedding of all requests is out-of-question, as this potentially
comes at a high cost and with long outage times. However, small
local reconfigurations may reduce the overall resource overhead and
improve the overall embedding substantially, or even make the
embedding possible at all.

To this end, we introduce matrices and constraints
that allow the specification of reconfiguration costs and enable the solver to weight them against the respective benefits.
Analogue to $new(u,v)$, we use the constant binary matrix $old(u,v)$ to
describe existing mappings, and
specify whether a virtual network element $u$ is currently mapped
to a substrate element $v$ ($old(u,v)=1$) or not ($old(u,v)=0$).

We account for the cost of migration in two respects: destination
independent cost factors are reflected in the constant penalty
matrix $penalty(u) = C_{\mrm{contract}}(u) + C_{\mrm{mgmt}}(u) + C_{\mrm{reconfig}}(u)$.
Destination dependent $C_{\mrm{transit}}$ and $C_{\mrm{adaptation}}$ cost factors to migrate
virtual network element $u$ from its current position
to substrate element $v$ are summed up in the constant matrix
$transit(u,v)$.

Node migration is typically more expensive relative to
link migration, as links do not involve state or bulk data transfers
but are rather re-instantiated.
As long as at least one end point of a modified link remains in place (i.e., the connected virtual node did not migrate), connectivity can be guaranteed by temporary redundant resource allocations. Costs can be reflected by the link's $transit(u,v)$ variables.
If all endpoints migrate simultaneously, at least one interconnecting segment or tunnel is required to allow for live-migrations.
As a simplification, we consider a scenario where every migrating host receives a temporary tunnel, as proposed in VROOM~\cite{vroom}, and where costs are added to the respective host's $transit(u,v)$ values.
Furthermore, we do not consider contract penalties for removed link segments, or migration-maximizing objective functions in this step.
We thus assume $penalty(u)=\epsilon$ for $\forall u\in NE_{\mrm{VL}}$ and an arbitrarily small $\epsilon>0$, unless stated otherwise.

\section{The Embedding Program}\label{sec:map:prog}

Based on the above ideas we next describe the MIP in details, see
Tables \ref{tab:sets}(sets), \ref{tab:constants}(constants), \ref{tab:variables}(variables) for a summary.
While we introduced most sets, variables, and constants above we will describe the remaining ones in support of specific objective functions, and proceed with an explanation of the constraints.

\begin{table*} [th]
\begin{center}
\begin{scriptsize}
\begin{tabular}{ | c | c |}
\hline
\multicolumn{2}{|c|}{Sets} \\
\hline
  $NE_\mrm{V}$ & Virtual Network Elements\\
  $NE_{\mrm{VN}}$ & Virtual Nodes\\
  $NE_{\mrm{VL}}$ & Virtual Links\\
  $NE_\mrm{S}$ & Substrate Network Elements\\
  $NE_{\mrm{SN}}$ & Substrate Nodes\\
  $NE_{\mrm{SL}}$ & Substrate Links\\
  $R_\mrm{V}$ & Set of Virtual Resource\\
  $R_\mrm{S}$ & Set of Substrate Resource\\
  $R_f: R_f \subset R_\mrm{V}$ & Set of Virtual Flow Resources\\
  $VT$ & Value Types\\
  $Fl(u)$ & Flows ((source,sink)-Tuples)\\
  \hline
\end{tabular}\label{sets}
\end{scriptsize}
\end{center}
\caption{Set definitions\label{tab:sets}}
\end{table*}


\begin{table*} [th]
\begin{center}
\begin{scriptsize}
\begin{tabular}{ | c | c | c | c |}
\hline
\multicolumn{2}{|c|}{Table~2: Constants} & \multicolumn{2}{|c|}{Range}\\
\hline
  $weight(u,v)$ & Resource Weight & $\forall u \in NE_\mrm{V}, v \in NE_\mrm{S}$ & $\in [0,1]$\\
  $penalty(u)$ & Migration Cost & $\forall u \in NE_\mrm{V}$ & $> 0$\\
  $transit(u,v)$ & Costs transferring $u$ resources to $v$ & $transit(u,v)$, $\forall u \in NE_{\mrm{V}}, v \in NE_\mrm{S}$ & $\geq 0$ \\
  $old(u,v)$ & Old Mapping & $\forall u \in NE_\mrm{V}$, $v \in NE_\mrm{S}$ & $\in \{0,1\}$\\
  $suit(u,v)$ & Suitable Mapping & $\forall u \in NE_\mrm{V}, v \in NE_\mrm{S}$ & $\in \{0,1\}$\\
  $cap_{r_\mrm{S}}(v)$ & Capacity of $v$ w.r.t.~$r_\mrm{S}$ & $\forall v \in NE_S, r_S \in R_S$ & $\geq 0$\\
  $cap_{r_\mrm{S}}(v,w)$ & Connection Capacity & $\forall (v,w) \in NE^2_\mrm{S}, r_\mrm{S} \in R_\mrm{S}$ & $\geq 0$\\
  $cap(r_\mrm{S})$ & Resource $r_\mrm{S}$ Capacity & $\forall r_\mrm{S} \in R_\mrm{S}$ & $\geq 0$\\
  $req(u,r_\mrm{V},s)$ & Resource Request & $\forall u \in NE_\mrm{V}, r_\mrm{V} \in R_\mrm{V},$ $ s \in VT$ & $\geq 0$\\
  $prop(r_\mrm{V},r_\mrm{S})$ & Scaling Factor & $\forall r_\mrm{V} \in R_\mrm{V}, r_\mrm{S} \in R_\mrm{S}$ & $\geq 0$\\
  $weight_{r_\mrm{S}}$ & Load Weight Factor& $\forall r_\mrm{S} \in R_\mrm{S}$ & $\in [0,1]$\\
  $c$ & \texttt{sum},\texttt{max} Load Priority Factor & & $\geq \sum_{r_\mrm{S} \in R_\mrm{S}}weight_{r_\mrm{S}}$\\
  $min\_alloc_{r_\mrm{V}}$ & Min. $r_\mrm{V}$ allocation unit & $\forall r_\mrm{V} \in R_\mrm{V}$ & $\geq 0$\\
  \hline
\end{tabular}\label{consts}
\end{scriptsize}
\end{center}
\caption{Constant definitions\label{tab:constants}}
\end{table*}


\begin{table*} [th]
\begin{center}
\begin{scriptsize}
\begin{tabular}{ | c | c | c | c |}
\hline
\multicolumn{2}{|c|}{Table~3: Variables} & \multicolumn{2}{|c|}{Range}\\
\hline
  $alloc_{r_\mrm{S}}(u,v,r_\mrm{V})$ & Allocated Resources & $\forall u \in NE_\mrm{V}, v \in NE_\mrm{S}, \forall r_\mrm{V} \in R_v ,r_\mrm{S} \in R_\mrm{S}$ & $\geq 0$\\
  $alloc_{r_\mrm{V}}(u,v)$ & Hosted Resources & $\forall u \in NE_\mrm{V}, v \in NE_\mrm{S}, \forall r_\mrm{V} \in R_\mrm{V}$ & $\geq 0$\\
  $new(u,v)$ & Mapping Matrix for Elements & $\forall u \in NE_\mrm{V}, v \in NE_\mrm{S}$ & $\in \{0,1\}$\\
  $new(f,v)$ & Mapping Matrix for Flows & $\forall f\in Fl(u), v \in NE_\mrm{S}, \forall u\in NE_{\mrm{VL}}$ & $\in \{0,1\}$\\
  $mig(u)$ & Migration Selector & $\forall u \in NE_\mrm{V}$ & $\in \{0,1\}$\\
  $flow_{r_\mrm{S}}(f,v,w,r_\mrm{V})$ & Allocated Resources for Flow & $\forall (v,w) \in NE_\mrm{S}^2, $ & $\geq 0$\\
	& & $\forall f \in Fl(u), r_\mrm{V}\in R_\mrm{V}, r_\mrm{S} \in R_\mrm{S}, \forall u \in NE_{\mrm{VL}}$ & \\
  $flow_{r_\mrm{V}}(f,v,w)$ & Hosted Resources for Flow & $\forall (v,w) \in NE_\mrm{S}^2, $ & $\geq 0$\\
	& & $\forall f \in Fl(u), r_\mrm{V} \in R_\mrm{V}, u \in NE_{\mrm{VL}}$ & \\
  $load(r_\mrm{S})$ & Load on Resource $r_\mrm{S}$ & $\forall r_\mrm{S} \in R_\mrm{S}$ & $\geq 0$\\
  $max\_load$ & Max Load over All $r_\mrm{S}$ & & $\geq 0$\\
  \hline
\end{tabular}\label{vars}
\end{scriptsize}
\caption{Variable definitions}\label{tab:variables}
\end{center}
\end{table*}


\begin{figure*} [tb!]   
\begin{scriptsize}
\noindent \textbf{\normalsize Nodes:}

\vspace*{1mm}
\begin{tabular}{lll}
  \texttt{map\_node}: & $\sum_{v \in NE_\mrm{S}} new(u,v)= 1$ & $\forall u \in NE_{\mrm{VN}}$ \\
  \texttt{set\_new}: & $alloc_{r_\mrm{S}}(u,v,r_\mrm{V})\leq cap_{r_\mrm{S}}(v)new(u,v)$ &
  $\forall u \in NE_{\mrm{VN}}, v \in NE_\mrm{S}, r_\mrm{V} \in R_\mrm{V}, r_\mrm{S} \in R_\mrm{S}$ \\
  \texttt{req\_min}: & $alloc_{r_\mrm{V}}(u,v)\geq new(u,v)req(u,r_\mrm{V},s)$ & $\forall u \in NE_{\mrm{VN}}, r_\mrm{V}\in R_\mrm{V}, r_\mrm{S} \in R_\mrm{S}, s = \texttt{minimum}$ \\   
  \texttt{req\_max}: & $alloc_{r_\mrm{V}}(u,v)\leq new(u,v)req(u,r_\mrm{V},s)$ & $\forall u \in NE_{\mrm{VN}}, r_\mrm{V}\in R_\mrm{V}, r_\mrm{S} \in R_\mrm{S},s = \texttt{maximum}$ \\
  \texttt{req\_con}: & $alloc_{r_\mrm{V}}(u,v)= new(u,v)req(u,r_\mrm{V},s)$ & $\forall u \in NE_{\mrm{VN}}, r_\mrm{V}\in R_\mrm{V}, r_\mrm{S} \in R_\mrm{S}, s = \texttt{constant}$
\end{tabular}

\vspace*{1mm}

\noindent \textbf{\normalsize Mapping:}

\vspace*{1mm}

\begin{tabular}{l l l}
  \texttt{relate\_V}: & $alloc_{r_\mrm{V}}(u,v)\geq {min\_alloc}_{r_\mrm{V}} \cdot
  new(u,v)$ & $\forall u \in NE_\mrm{V}, v \in NE_\mrm{S}, r_\mrm{V} \in R_\mrm{V}$ \\
  \texttt{allowed}: & $suit(u,v)\geq new(u,v)$ & $\forall u \in NE_\mrm{V}, v \in NE_\mrm{S}$ \\
  \texttt{ne\_capacity}: & $\sum_{u \in NE_\mrm{V}}\sum_{r_\mrm{V} \in R_\mrm{V}}alloc_{r_\mrm{S}}(u,v,r_\mrm{V})\leq cap_{r_\mrm{S}}(v)$ & $\forall v \in NE_\mrm{S}, r_\mrm{S} \in
  R_\mrm{S}$ \\
  \texttt{capacity}: & $\sum_{v \in NE_\mrm{S}}\sum_{u \in NE_\mrm{V}}\sum_{r_\mrm{V} \in R_\mrm{V}}alloc_{r_\mrm{S}}(u,v,r_\mrm{V})\leq cap(r_\mrm{S})$ & $\forall r_\mrm{S} \in R_\mrm{S}$ \\
  \texttt{load}: & $weight_{r_\mrm{S}}/cap(r_\mrm{S}) \cdot $ & $\forall r_\mrm{S} \in R_\mrm{S}$ \\
   & $\sum_{v \in NE_\mrm{S}}\sum_{u \in NE_\mrm{V}}\sum_{r_\mrm{V} \in R_\mrm{V}}alloc_{r_\mrm{S}}(u,v,r_\mrm{V})\leq load(r_\mrm{S})$ & \\
  \texttt{max\_load}: & $load(r_\mrm{S})\leq max\_load$ & $\forall r_\mrm{S} \in R_\mrm{S}$
\end{tabular}

\vspace*{1mm}

\noindent \textbf{\normalsize Resource-Variable Relation:}

\vspace*{1mm}

\begin{tabular}{l l l}
  \texttt{resource}: & $\sum_{r_\mrm{S} \in R_\mrm{S}}prop(r_\mrm{V},r_\mrm{S})alloc_{r_\mrm{S}}(u,v,r_\mrm{V})= alloc_{r_\mrm{V}}(u,v)$
    & $\forall u \in NE_\mrm{V},v \in NE_\mrm{S}, r_\mrm{V} \in R_\mrm{V}$ \\
  \texttt{flow\_res}: & $\sum_{r_\mrm{S} \in R_\mrm{S}}prop(r_\mrm{V},r_\mrm{S})flow_{r_\mrm{S}}(f,v,w,r_\mrm{V})= flow_{r_\mrm{V}}(f,v,w)$
    & $\forall f \in Fl(u), (v,w) \in NE_\mrm{S}^2, r_\mrm{V} \in R_f,$\\
	& & $\forall u \in NE_{\mrm{VL}}$\\
\end{tabular}

\vspace*{1mm}

\noindent \textbf{\normalsize Links:}

\vspace*{1mm}

\begin{tabular}{l l l}
  \texttt{map\_link}: & $\sum_{v \in NE_\mrm{S}} new(u,v)\geq 1$
  & $\forall u \in NE_{\mrm{VL}}$ \\
  \texttt{map\_src}: & $new(u,v)\geq new(q_f,v)$ & $\forall f \in Fl(u), v \in NE_\mrm{S}, q_f$ source of $f ; \forall u \in NE_{\mrm{VL}}$  \\
  \texttt{map\_sink}: & $new(u,v)\geq new(d_f,v)$ & $\forall f \in Fl(u), v \in NE_\mrm{S}, d_f$ sink of $f ; \forall u \in NE_{\mrm{VL}}$ \\  
\end{tabular}

\vspace*{1mm}
\begin{tabular}{ll}
  \texttt{req\_fmin}: & $\sum_{w \in NE_\mrm{S}}(flow_{r_\mrm{V}}(f,v,w)-flow_{r_\mrm{V}}(f,w,v))$ $~~\geq new(q_f,v)req(u,r_\mrm{V},s) -
new(d_f,v)\infty$\\
  & $\forall f \in Fl(u), v \in NE_\mrm{S}, r_\mrm{V} \in R_f ; \forall u \in NE_{\mrm{VL}}, s = \texttt{minimum}$ \\
  \texttt{req\_fmax}: & $\sum_{w \in NE_\mrm{S}}(flow_{r_\mrm{V}}(f,v,w)-flow_{r_\mrm{V}}(f,w,v))$ $ ~~\leq
new(q_f,v)req(u,r_\mrm{V},s) + new(d_f,v)\infty$ \\
  & $\forall f \in Fl(u), v \in NE_\mrm{S}, r_\mrm{V} \in R_f ; \forall u \in NE_{\mrm{VL}}, s = \texttt{maximum}$ \\
  \texttt{req\_fconst}: & $\sum_{w \in NE_\mrm{S}}(flow_{r_\mrm{V}}(f,v,w)-flow_{r_\mrm{V}}(f,w,v))$ $~~ = new(q_f,v)req(u,r_\mrm{V},s)-
new(d_f,v)req(u,r_\mrm{V},s)$
  \\ & $\forall f \in Fl(u), v \in NE_\mrm{S}, r_\mrm{V} \in R_f ; \forall u \in NE_{\mrm{VL}}, s = \texttt{constant}$\\
\end{tabular}

\vspace*{1mm}

\noindent \textbf{\normalsize Link Allocation:}

\vspace{1mm}

\begin{tabular}{lll}
  \texttt{exp\_out}: & $\sum_{w \in NE_\mrm{S}} flow_{r_\mrm{S}}(f,v,w,r_\mrm{V})\leq alloc_{r_\mrm{S}}(u,v,r_\mrm{V})$ & $\forall f \in Fl(u), v \in NE_\mrm{S}, r_\mrm{V} \in R_f,$\\
	& & $r_\mrm{S} \in R_\mrm{S}, \forall u \in NE_{\mrm{VL}}$ \\
  \texttt{exp\_in}: & $\sum_{w \in NE_\mrm{S}} flow_{r_\mrm{S}}(f,w,v,r_\mrm{V})\leq alloc_{r_\mrm{S}}(u,v,r_\mrm{V})$ & $\forall f \in Fl(u), v \in NE_\mrm{S}, r_\mrm{V} \in R_f,$\\
	& & $r_\mrm{S} \in R_\mrm{S}, \forall u \in NE_{\mrm{VL}}$ \\   
  \texttt{direction}: & $flow_{r_\mrm{S}}(f,v,w,r_\mrm{V})\leq new(u,v)cap_{r_\mrm{S}}(v,w)$ & $\forall f \in Fl(u), (v,w) \in NE_\mrm{S}^2,$\\
	& & $r_\mrm{V} \in R_f, r_\mrm{S} \in R_\mrm{S},  \forall u \in NE_{\mrm{VL}}$ \\
  \texttt{relate\_f}: & $\sum_{w \in NE_\mrm{S}} flow_{r_\mrm{S}}(f,v,w,r_\mrm{V})+ flow_{r_\mrm{S}}(f,w,v,r_\mrm{V})\geq  new(u,v)$ & $\forall f \in Fl(u), \forall u \in NE_{\mrm{VL}},$\\
	& & $v \in NE_\mrm{S}, r_\mrm{V} \in R_f, r_\mrm{S} \in R_\mrm{S}$ \\
\end{tabular}

\vspace*{1mm}

\noindent \textbf{\normalsize Migration:}

\vspace*{1mm}

\begin{tabular}{l l l}
  \texttt{new}: & $\sum_{v \in NE_\mrm{S}} old(u,v)\geq mig(u)$ & $\forall u \in NE_\mrm{V}$ \\
  \texttt{migrated}: & $old(u,v)-new(u,v)\leq mig(u)$ & $\forall u \in NE_\mrm{V}, v \in NE_\mrm{S}$
\end{tabular}
\end{scriptsize}
\begin{center}
\caption{Embedding constraints for linear Mixed Integer Program.
Explanations are given in the text.}\label{fig:constraints}
\end{center}
\end{figure*}

\subsection{Objective Function}\label{ssec:mip:prog:objective}

How is an optimal embedding of a \CloudNet\ on a
set of resources in the substrate network characterized?
The answer depends on the goals of the mapping entity, and also
relies crucially on the predictability of future resource requests.
Even with good predictions, an optimal solution found at time $t_0$
may be suboptimal upon the arrival of the next request at some time
$t>t_0$.

We hence do not propose any specific objective functions here (recall that one advantage of our mathematical programming approach is the ease of exchanging different objective functions with only limited implementation effort) but just consider two canonical examples:
minimizing resource usage (to ensure localizing embeddings) and load balancing (by spreading the embeddings as much as possible).

%

The minimization of the amount of substrate resources used for \CloudNet\ allocations is a natural objective that maximizes the chances to be able to embed also future requests, to save energy by switching off unused hardware, or to perform maintenance work. The objective function also used for our experiments hence balances resource usage and migration cost:

\begin{footnotesize}
$$\sum_{u \in NE_\mrm{V}} \sum_{v \in NE_\mrm{S}} \sum_{r_\mrm{S} \in R_\mrm{S}}
weight(u,v)\cdot alloc_{r_\mrm{S}}(u,v,r_\mrm{V})$$
$$
+ \sum_{u \in NE_\mrm{V}}\left( penalty(u)\cdot mig(u) + \sum_{v \in
NE_\mrm{S}} transit(u,v)\cdot new(u,v) \right)$$
\end{footnotesize}

Alternatively, in our prototype, we also employ an objective function that seeks to
distribute the load equally among all network elements to minimize
peak loads and congestion.
Such an objective function may make sense, if requests are likely to involve placement constraints, or if resource guarantees allow for usage spikes:

\begin{footnotesize}
$$c\cdot max\_load + \sum_{r_\mrm{S} \in R_\mrm{S}} load(r_\mrm{S})$$
$$+
\sum_{u \in NE_\mrm{V}} \left( penalty(u)\cdot mig(u) + \sum_{v \in
NE_\mrm{S}} transit(u,v)\cdot new(u,v) \right)$$
\end{footnotesize}

To this end, we extend the program by the
$load(r_\mrm{S})$ matrix capturing the individual substrate resource loads
(needed in the objective function for efficient allocation).
$max\_load$
denotes the maximal load over all resources and is defined in the
constraints of the MIP.
This dual load approach is required to compensate for variation in availability of different resources:
minimizing only $max\_load$ would optimize only the  scarcest resource and hence leave overly high slack in other resource allocations.
Minimizing individual $load(r_\mrm{S})$ avoids unnecessary resource allocations, but (again numerically) overrules $max\_load$ as a prime factor.
Therefore, the constant factor $c$ is required to balance between overall and individual load.

\subsection{Constraints}\label{ssec:mip:prog:constraints}

The embedding must fulfill various type, capacity, and other
consistency constraints, see Figure~\ref{fig:constraints} for a
complete and formal constraint list.

\begin{description*}
\item \textbf{Nodes:} This constraints category is used to ensure that each \CloudNet\
  node is mapped to an appropriate substrate node. In contrast to
  links, we do not map nodes to multiple substrate elements, and hence
  Constraint \texttt{map\_node} is necessary to guarantee a unique mapping
  location. At the location where the node is mapped (and only there!),
  resource requirements must be fulfilled (Constraint
  \texttt{set\_new}). Depending on the substrate resource type
  (\texttt{minimum}, \texttt{maximum}, or \texttt{constant}), the resource
  constraints are imposed in a different manner (Constraints
  \texttt{req\_min}, \texttt{req\_max}, \texttt{req\_con}).

\item \textbf{Mapping:} The mapping constraints ensure that the substrate element has
  sufficient capacity (Constraint \texttt{ne\_capacity}) allocated.
  If resources are shared amongst substrate elements, we need to check against the capacity of the resource itself (Constraint \texttt{capacity}).
  In order to limit link splitups, we set a minimal resource allocation unit (Constraint \texttt{relate\_V}).
  Moreover, virtual elements hosted must be of the correct type
  (Constraint \texttt{allowed}).
  Constraint \texttt{load} and Constraint
  \texttt{max\_load} define the load of a resource (i.e. the fraction of its capacity used)
  and the maximum of all individual resourc
  loads, respectively.

\item \textbf{Resource-Variable Relation:} This set of constraints deals with the
  relation between the resource types $r_\mrm{S}$ that host resources of type $r_\mrm{V}$.
  In our mathematical program, we assume a linear relation, which is given by
  the constant factor $prop(r_\mrm{V},r_\mrm{S})$ (Constraints \texttt{resource} and \texttt{flow\_res}).

\item \textbf{Links:} Mapping links is similar to mapping nodes, and hence, several
  constraints apply also to links.
  However, in contrast to nodes, links may be mapped to more than one substrate element (as one or several paths).
  Shared communication channels need to allocate resources to satisfy their requirements with respect to every virtual node pair connected.
  In order to calculate allocations in this case, links are expanded into a set of flows, as described earlier.
  Clearly, each virtual link must be mapped to at least one substrate element
  (Constraint~\texttt{map\_link}).
  Sources and sinks of the expanded flows definitely are part of this mapping
  (Constraints~\texttt{map\_src} and~\texttt{map\_sink}).
  Note that this allows to find a valid mapping even for pure local links, i.e. if all virtual nodes are mapped to a single substrate node.
  As a simplification, we assume that pure local links require only nominal resources, considering only resource allocation corresponding to $min\_alloc_{r_\mrm{V}}$\footnote{This can be extended trivially by adding a variant of constraints $req\_*$ for links $u$, where $new(u,v)$ is replaced by $new(u',v)$ for all virtual nodes $u'$ connected to $u$}.

  The multi-path propagation of each flow $f$ must satisfy flow preservation,
  except for the source and sink element. The constraint depends on the
  value type (\texttt{minimum}, \texttt{maximum}, or \texttt{constant}): In
  case of a minimum type, the net flow of a given resource type must be a
  least the requested resources at the source and preserved otherwise. If the
  substrate element is the sink, the flow preservation invariant is suspended
  and the constraint becomes fulfilled trivially. To implement a corresponding
  selector, multiplication by a sufficiently large number (e.g., slightly larger than
  the maximal amount of involved resources, here simply represented by $\infty$) is used in the subtrahend.
  This yields the desired tautology (see Constraint~\texttt{req\_fmin}). The
  Constraints~\texttt{req\_fmax} and~\texttt{req\_fconst} are defined analogously.
 Note however that it is not possible to mathematically strictly ensure maximum or constant bandwidth in combination with half-duplex links.

\item \textbf{Link Allocation:}
  The $r_\mrm{S}$ allocated for a virtual $u$ on a substrate element $v$ is the maximum of the $r_\mrm{S}$ required for every single of $u$'s flows.
  Constraints~\texttt{exp\_out} and~\texttt{exp\_in}
  ensure that these resources are allocated on sources and destinations of the respective flows.
  Constraint~\texttt{direction} enforces direction specific capacity constraints on full-suplex substrate resources.

\item \textbf{Migration:} Our program allows us to migrate already embedded cloud network
  elements to new locations, if the reconfiguration costs are amortized by the
  more efficient embedding.
  The migration constraints set the migration flag
  $mig(u)$ if\footnote{and only if, whenever migration costs are relevant - i.e., $> 0$, and minimized in the objective function}
  the mapped element is not new (Constraint
  \texttt{new}) and was previously embedded at a different location, where it was removed
  (Constraint \texttt{migrated}).
\end{description*}

\section{Experiments} \label{sec:experiments}

This section reports on our experience with the prototype
implementation. We conducted experiments for the out-sourcing/cloud
(OC) and the data center (DC) scenarios; the virtual private
networks (VPN) use case is studied under the out-sourcing scenario
by setting the placement freedom to zero (i.e., all virtual cloud
nodes have a fixed location).

In order to model the physical substrate network, we extracted
\emph{Rocketfuel topologies}~\cite{rocketfuel}. Connected subsets of
these graphs are also used to describe the topology of the
\CloudNet\ requests. For the OC use case, the virtual cloud nodes of
the \CloudNets\ are partitioned into freely allocatable cloud
resources (CR) and fixed access points (AP) (e.g., connection points
to corporate subnets of the requesting entity); in the DC use case
all virtual cloud nodes exhibit full placement flexibility. Unless
stated otherwise, substrate network elements feature capacity for
fifteen virtual network elements, no placement preferences are given
(with destination independent migration penalties for all node pairs
and with unit weights)
and the
embeddings are optimized with the maximal load minimization
objective function. As a solver, IBM's standard \texttt{CPLEX}
software is used in deterministic mode with a limit of six
concurrent threads on a 8-core Xeon server running at 2.5GHz.

\subsection{Out-sourcing Scenario}

In the out-sourcing scenario, the virtual cloud nodes fall into two
categories: a set of fixed APs, and a set of freely placeable CRs.
Concretely, for each \CloudNet\ we chose (uniformly at random)
between one and three flexible cloud nodes and between one and seven
fixed access nodes. We refer to the percentage of flexible nodes in
the \CloudNet\ by the variable $freedom\in[0,1]$; note that
$freedom=0$ is our VPN scenario. For our experiments, we use a
substrate network of twenty-five nodes, and we iteratively place
incoming \CloudNet\ requests. Evaluations are repeated ten times,
and all \CloudNet\ requests are accepted as long as resources are
available. We study scenarios with and without migration.

\begin{figure*} [tb]
\begin{center}
\vspace*{-.75cm}
\includegraphics[width=0.9\columnwidth]{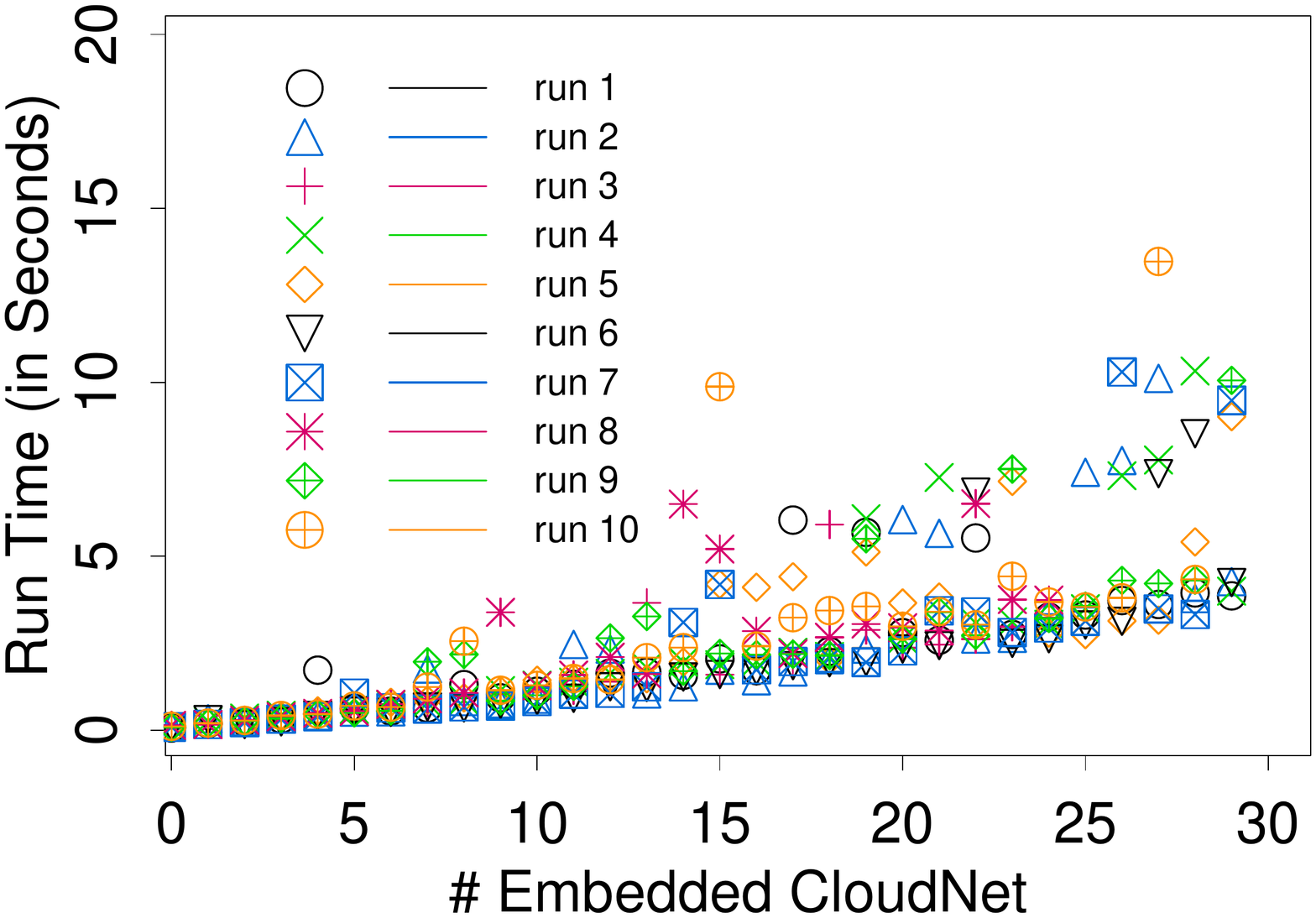}
\includegraphics[height=0.9\columnwidth,angle=90]{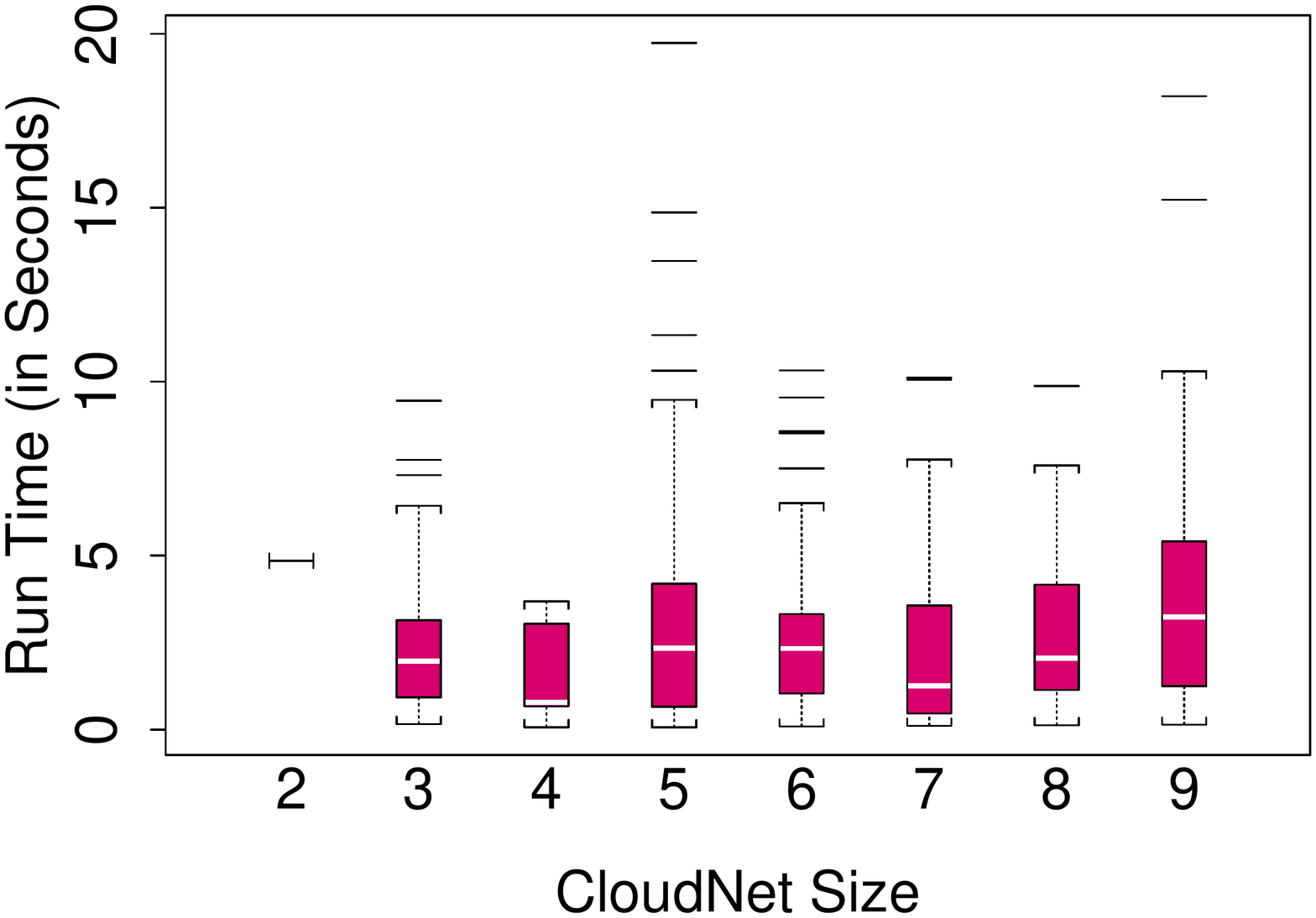}
\caption{Run time per embedded \CloudNet\ without migration for multiple
  runs (left: time series, right: boxplot).}\label{fig:exp-withoutmigration}
\end{center}
\end{figure*}

Figure~\ref{fig:exp-withoutmigration} shows the runtime (real time,
in seconds) required to embed \CloudNets\ iteratively (one after the
other, sorted on the x-axis) in a scenario without migrations. There
are several takeaways from these experiments: First, we observe that
the embedding times are small (never exceeding 14 seconds). Moreover,
depending on the load on the substrate network (the number of
already embedded \CloudNets), the runtime increases slightly. The
data also exhibits a relatively high variance, which can be
explained by the randomized nature of the to be embedded \CloudNets\ (in
terms of size and nature).

\begin{figure} [tb]
\vspace*{-.75cm}
\begin{center}
\includegraphics[width=0.9\columnwidth]{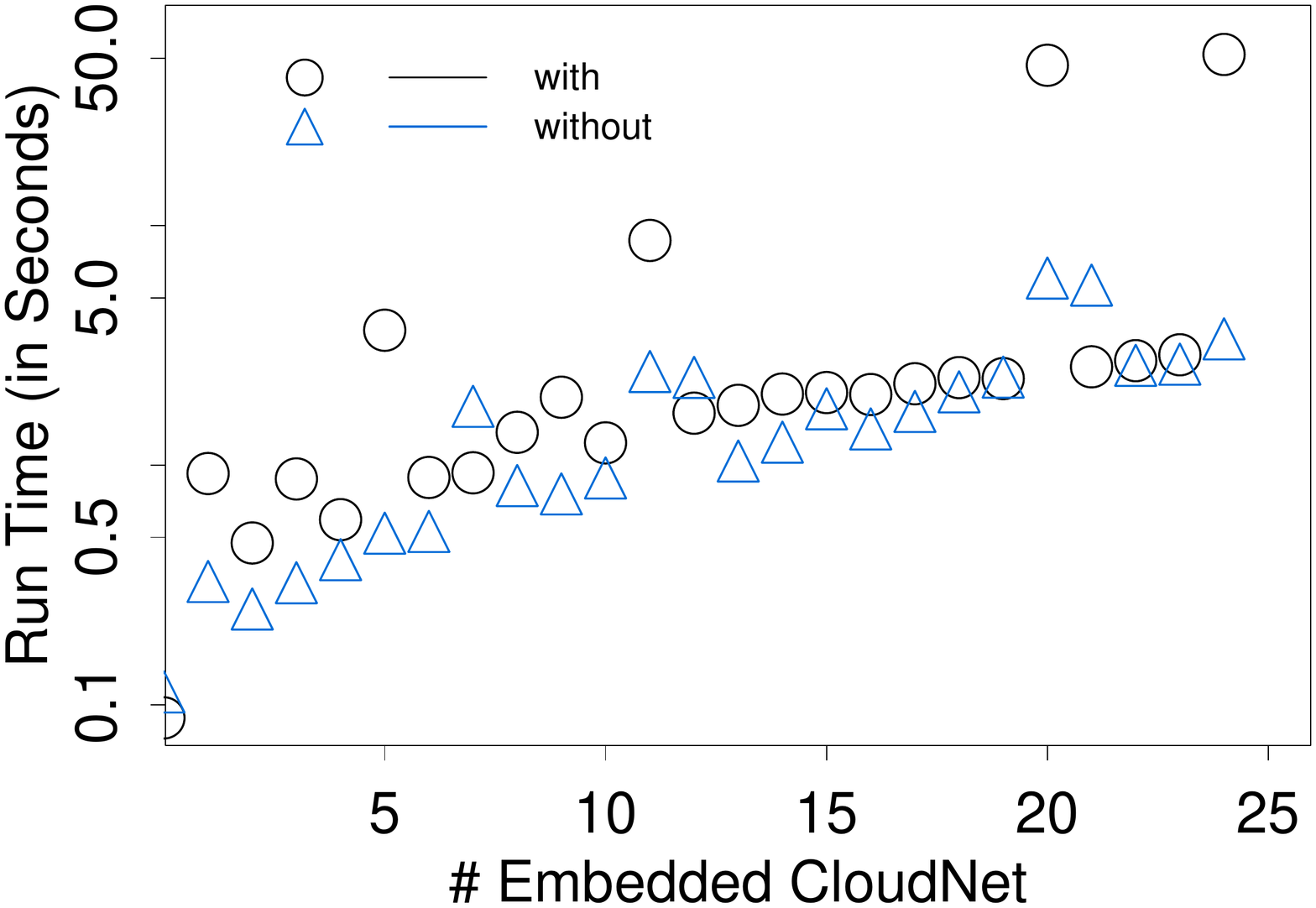}\\
\caption{Run time (log scale) per embedded \CloudNet\ with and without
  migration (sample run).}\label{fig:exp-withandwithoutmigration}
\end{center}
\end{figure}

The run times generally increase if we enable the option
to migrate, see Figure~\ref{fig:exp-withandwithoutmigration}, although there
are instances where the results are comparable. This is to be
expected as migration increases flexibility and therefore the complexity of
the MIP.
An interesting feature of integrating migration support is that we can at any
time check if a subset of the resources, e.g., half of the network is
sufficient to fulfill the demand. In the above cases such a run takes on
average 2.73 seconds with a standard deviation of 0.42 seconds.
%

\begin{figure} [tb]
\vspace*{-.75cm}
\begin{center}
\includegraphics[width=0.9\columnwidth]{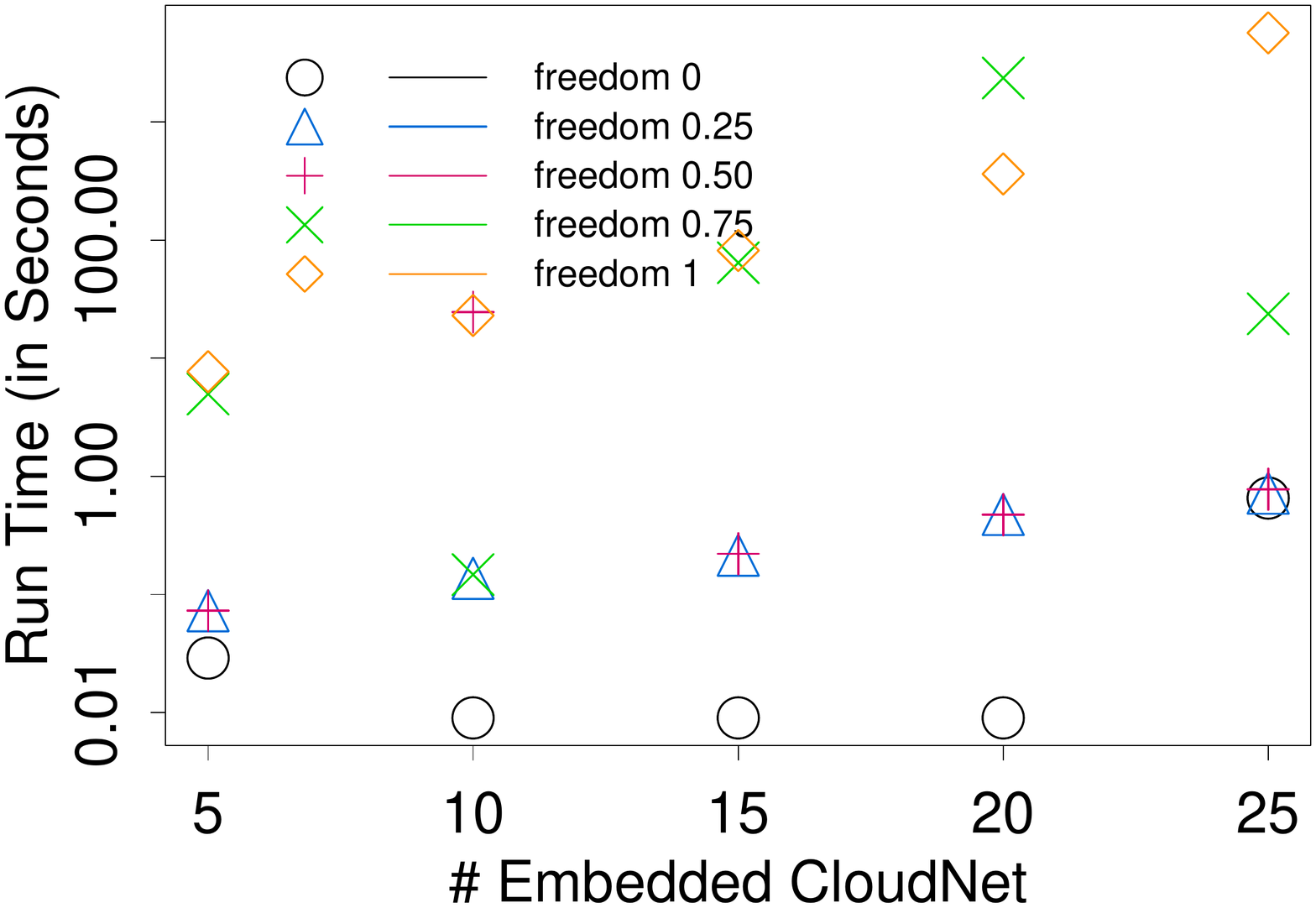}\\
\caption{Run time (log scale) per embedded \CloudNet\ for different
  degrees of freedom (sample runs). \texttt{CPLEX} is
used in deterministic mode yielding constant
run times.}\label{fig:exp-freedom}
\end{center}
\end{figure}

As the above experiment suggests, the main
parameter that determines the time complexity of the embeddings is
the freedom of the node placement. We conducted on a series of
experiments where the \CloudNet\ size and the proportion of CRs,
i.e., the variable $freedom$, varies. The findings are summarized in
Figure~\ref{fig:exp-freedom} which confirms this dependency.
Interestingly, despite the flexibility of the \CloudNets\ and the
existing load on the substrate, the run times are still in the range
of several minutes.

We can conclude that although the option to migrate and the
placement flexibility effect the execution times, optimal solutions
for relatively large problems are feasible and can be computed in
reasonable time. Moreover, as the run times without migration support
are lower than their counterparts, hybrid designs, where incoming
\CloudNets\ are first placed ad-hoc, and persisting ones are
optimized regularly (by an offline, background process) are
attractive.

\subsection{Data Centers}

The data center use case exhibits the highest flexibility and hence
constitutes, in some sense, the ``worst case'' scenario (in terms of
embedding complexity). In order to quantify the impact of the
substrate network size, we calculated mappings for a single
twenty-five CR \CloudNet\ on substrates of different sizes. (The complexity is
similar to experiments with multiple \CloudNets\ using the same amount of
resources.) In one
set of experiments, we calculated the optimum, in another we
emphasized feasibility. For the latter experiments, we turned off the
multi-threading and parallelization features of \texttt{CPLEX}.

\begin{figure} [tb]
\vspace*{-.75cm}
\begin{center}
\includegraphics[width=0.9\columnwidth]{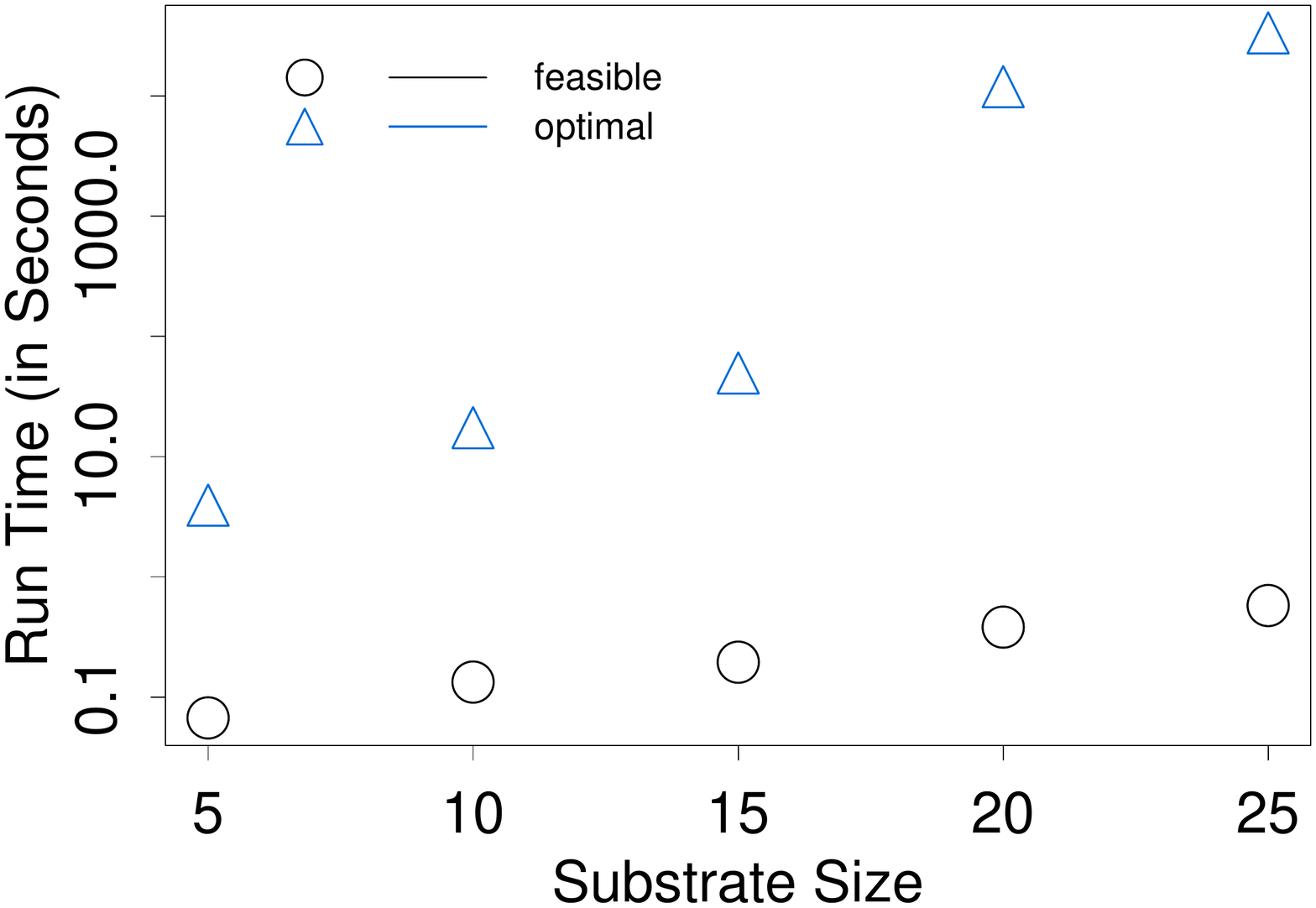}\\
\caption{Run time (log scale) vs.\ substrate size for full
  degrees of freedom---optimal solution vs.\ feasible solution (sample
  runs).}\label{fig:exp-substratesize}
\end{center}
\end{figure}

Figure~\ref{fig:exp-substratesize} studies the price of optimality:
the comparison of the run times for optimal and feasible (i.e., first
possible) \CloudNet\ embeddings shows that while the performance of
both depends on the substrate size, optimal solutions may result in an order
of magnitude higher run times. For a substrate network of around
twenty nodes, we can still expect an optimal solution within hours.
Recall, this is a worst case scenario as it offers full flexibility.

\begin{figure} [tb]
\vspace*{-.75cm}
\begin{center}
\includegraphics[width=0.9\columnwidth]{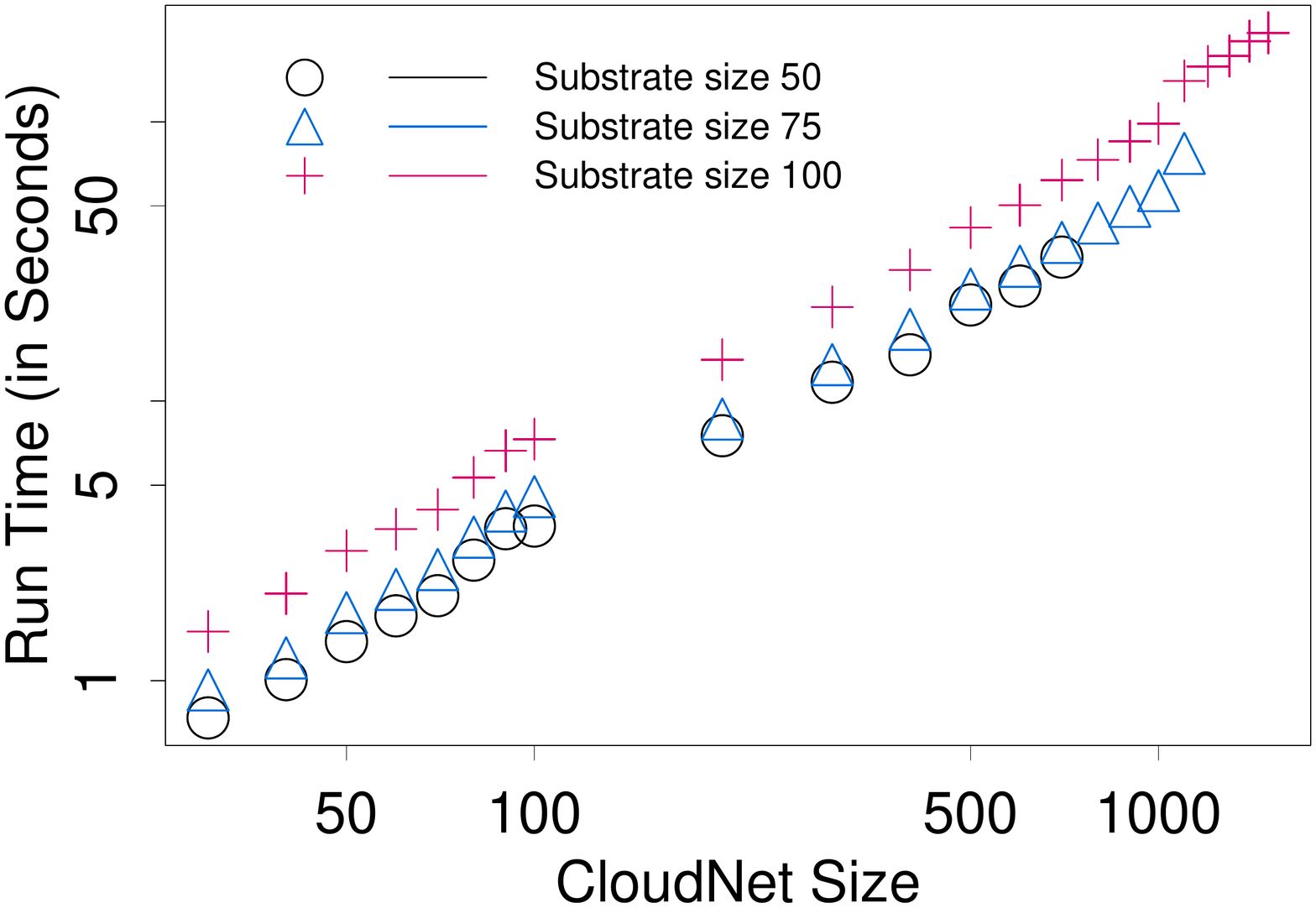}\\ %
\caption{Run time (log scale) vs.\ \CloudNet size (log scale) for different
  substrate size (sample runs).}\label{fig:exp-cloudnetsize}
\end{center}
\end{figure}

Finally, we examine feasibility in detail and report on an
experiment studying the run times as a function of the \CloudNet\
size (cf~Figure~\ref{fig:exp-cloudnetsize}): the loglog plot
indicates that the runtime grows linearly for larger network sizes.
Even \CloudNets\ with thousands of elements can be embedded within
minutes in a substrate of almost one hundred nodes.

As a final remark, note that although our run times for the data
center embeddings are higher than in the out-sourcing use case, the
conclusion that \CloudNet\ optimizations are time-consuming in data
centers may be wrong. Our experiments are overly conservative in the
sense that in reality, we expect that \CloudNet\ requests for real
data centers to be homogeneous, and are issued to, e.g., computing
grid environments which are typically likely to be homogeneous as
well. This homogeneity facilitates a range of optimizations, e.g.,
by aggregating entities into larger meta-nodes; this reduces the
number of elements (i.e., variables) in the problem.

\section{Related Work} \label{sec:relwork}

There has been a significant interest in virtual networks over the
last years.The reader is
referred to the recent surveys~\cite{virsurvey} and~\cite{chalsurvey}. The
work described here is conducted in the context of our network
virtualization project where we develop a prototype implementation.

Network embedding problems have already been studied in various
settings (even in the context of circuit-design, which is however
quite different and not discussed further here). Note that the
virtual network embedding problem is different from classic
\emph{VPN embedding} or \emph{multi-flow} problems in the sense that
the node placement is not given but subject to optimization as well.
This additional degree of freedom renders the problem more complex.
Indeed, many variants already of much simpler virtual network
embedding problems are computationally hard: Even if all virtual
network requests are given in advance, the offline optimization
problem with constraints on virtual nodes and virtual links can be
reduced to the NP-hard \emph{multi-way separator problem} (e.g.,
see~\cite{nphard} for a survey). Thus, there is a large body of
literature on \emph{heuristic solutions}: For example, Fan and
Ammar~\cite{ammar} study dynamic re-configurable topologies to
accommodate communication requirements that vary over time, Zhu and
Ammar~\cite{zhu06} consider virtual network assignment problems with
and without reconfiguration but only for bandwidth constraints,
Ricci et al.~\cite{simannealing} pursue a simulated annealing
approach, and Lu and Turner~\cite{turner} seek to find the best
topology in a family of backbone-star topologies. Many approaches in
the literature fail to exploit the flexibility to embed virtual
nodes and links simultaneously and solve the two mappings
sequentially (e.g.,~\cite{pb-embed}), which entails a loss of
efficiency~\cite{isosub}. To deal with the computational hardness,
Yu et al.~\cite{rethinking} advocate to rethink the design of the
substrate network to simplify the embedding, e.g., by allowing to
split a virtual link over multiple paths and perform periodic path
migrations. The focus of the work by Butt et
al.~\cite{topoawareness} is on re-optimization mechanisms that
ameliorate the performance of the previous virtual network embedding
algorithms in terms of acceptance ratio and load balancing; their
algorithm is able to prioritize resources and is evaluated by
simulations. Virtual network embeddings have also been
studied from a distributed computing
point-of-view~\cite{distmapping}.

In contrast to the literature reviewed above, our work seeks to
combine virtual networks with storage and computation to enable
virtual cloud networks, and puts an emphasis on \emph{generality} and
\emph{quality} of the embeddings. (Of course, our mathematical
program can also be solved heuristically, e.g., for ad-hoc
placements.) We believe that the mathematical programming approach
we pursued has many advantages, as it allows for a simple
replacement of the objective function, and as state-of-the-art
and optimized solvers can be used to find not only optimal but also
approximate or heuristic solutions. (There is no need to reinvent,
e.g., new pruning heuristics for each embedding problem variant;
often such heuristics are also unlikely to be faster than the
sophisticated algorithms incorporated into \texttt{CPLEX} or
\texttt{lpsolve}.) We are only aware of two embedding problems
related to virtual networks for which a mathematical program exists:
Kumar et al.~\cite{hose} describe an approach to solve a Virtual
Private Network tree computation problem for bandwidth provisioning;
flexible virtual node placements are not possible. Chowdhury et
al.~\cite{infocom2009} present an integer embedding program and
pursue a relaxation strategy, applying randomized and deterministic
rounding to find approximate solutions. The presented graph
extension approach supports exact solutions for placements where
interconnected virtual nodes do not share candidate substrate nodes,
as it would add bogus resources otherwise.
Hajjat et al. \cite{clwardsigc11} calculate
reconfigurations in the context of enterprise applications under the
constraint of communication costs after migration. Costs of the
migration itself are not considered.

To the best of our knowledge, there is no algorithm to embed
\CloudNet\ like networks in a manner whose generality and
flexibility is close to ours. In particular, none of the solutions
above can handle all the heterogeneous links occurring in practice
and map, e.g., a (wireless) broadcast link onto a set of asymmetric
and full-duplex links; besides the virtual links, also the
expressiveness of the node mapping is restricted, and we are not
aware of any algorithm which e.g., allows to capture loads induced
due to packet rates of the flows in a \CloudNet; finally, we believe
that the support of cost-aware migration is crucial, as the
dynamical aspects lie at the heart of network virtualization.

\section{Conclusion}\label{sec:conclusion}

We have proposed an integrative and flexible approach to realize
\CloudNets\ by jointly considering node and link placement on
heterogeneous resources. Moreover, our algorithm considers ``the use
of migration'' as an important primitive and thus allows the
operator to study, i.e., the trade-off between the gains (e.g., in
terms of resource savings or QoS) that can be obtained from
migrating existing \CloudNets\ to different locations, and the
corresponding migration cost. We find that joint optimal embeddings
of long-lived \CloudNets\ are feasible for moderate size networks, especially
in a hierarchical management hierarchy as we envision it in our
federated prototype architecture and
implementation~\cite{visa09virtu}. Moreover, we believe that the
computations can be further sped up by optimizing the solver. In
future work, we plan to continue the study of the quality of our
embeddings over time, i.e., to devise online algorithms that (in
contrast to VPN embedding approaches~\cite{moti}) exploit the
placement flexibility in a competitive manner.

{\footnotesize
\bibliographystyle{abbrv}
\bibliography{prototyp}
}

\end{document}